\documentclass[aps,prb,twocolumn,superscriptaddress]{revtex4}

\usepackage{amssymb}

\usepackage{color}

\usepackage{graphicx}

\begin{document}

\bibliographystyle{naturemag}

\title{Built-in and induced polarization across LaAlO$_3$/SrTiO$_3$ heterojunctions}

%The ionic potential and induced dipoles at polar interfaces

%Direct measure of the ionic potential in polar thin films

\author{Guneeta Singh-Bhalla}
\email[Corresponding author:~]{guneeta@berkeley.edu}
\affiliation{Department of Physics, University of California, Berkeley, California 94720, USA}
\affiliation{Materials Science Division, Lawrence Berkeley National Laboratory, Berkeley, California 94720, USA}
\affiliation{Department of Advanced Materials Science, University of Tokyo, Kashiwa, Chiba 277, Japan}
\affiliation{Department of Physics, University of Florida, Gainesville, Florida 32611, USA}

\author{Christopher Bell}
\affiliation{Department of Advanced Materials Science, University of Tokyo, Kashiwa, Chiba 277, Japan}
\affiliation{Japan Science and Technology Agency, Kawaguchi, 332-0012, Japan}

\author{Jayakanth Ravichandran}
\affiliation{Applied Science and Technology Graduate Group, University of California, Berkeley, California 94720, USA}

\author{Wolter Siemons}
\affiliation{Department of Physics, University of California, Berkeley, California 94720, USA}

\author{Yasuyuki Hikita}
\affiliation{Department of Advanced Materials Science, University of Tokyo, Kashiwa, Chiba 277, Japan}

\author{Sayeef Salahuddin}
\affiliation{Department of Electrical Engineering and Computer Science, University of California, Berkeley, California 94720, USA}

\author{Arthur F. Hebard}
\affiliation{Department of Physics, University of Florida, Gainesville, Florida 32611, USA}

\author{Harold Y. Hwang}
\affiliation{Department of Advanced Materials Science, University of Tokyo, Kashiwa, Chiba 277, Japan}
\affiliation{Japan Science and Technology Agency, Kawaguchi, 332-0012, Japan}

\author{Ramamoorthy Ramesh}
\affiliation{Department of Physics, University of California, Berkeley, California 94720, USA}
\affiliation{Materials Science Division, Lawrence Berkeley National Laboratory, Berkeley, California 94720, USA}

\date{\today}

\newcommand{\lao}{{LaAlO$_3$}}
\newcommand{\sto}{{SrTiO$_3$}}

\begin{abstract}
Ionic crystals terminated at oppositely charged polar surfaces are inherently unstable and expected to undergo surface reconstructions to maintain electrostatic stability.  Essentially, an electric field that arises between oppositely charged atomic planes gives rise to a built-in potential that diverges with thickness. In ultra thin film form however the polar crystals are expected to remain stable without necessitating surface reconstructions, yet the built-in potential has eluded observation.  Here we present evidence of a built-in potential across polar \lao ~thin films grown on \sto ~substrates, a system well known for the electron gas that forms at the interface.  By performing electron tunneling measurements between the electron gas and a metallic gate on \lao ~we measure a built-in electric field across \lao ~of 93 meV/\AA.   Additionally, capacitance measurements reveal the presence of an induced dipole moment near the interface in \sto, illuminating a unique property of \sto ~substrates.  We forsee use of the ionic built-in potential as an additional tuning parameter in both existing and novel device architectures, especially as atomic control of oxide interfaces gains widespread momentum.
\end{abstract}


\maketitle

The creation of oxide thin films with polar surfaces has only recently become possible with the advent of termination control on SrTiO$_3$ (001) substrates.~\cite{kawasaki1994,koster1998}  \lao ~thin films grown on singly terminated \sto ~surfaces comprise negatively charged AlO$_2$ and positively charged LaO end planes and are expected to be polar in the ionic limit.~\cite{nakagawa2006,Tasker1979,Goniakowski2008}  When at least four unit cells (u.c.) of \lao ~are deposited on TiO$_2$ terminated \sto, an electron gas forms near the interface in \sto.~\cite{ohtomo2004,Thiel2006,segal2009}  It is often hypothesized that at the critical thickness of four u.c.\ the potential across \lao ~exceeds the band gap of \sto ~and electrons tunnel from the valence band of \lao ~to the \sto ~potential well, completely diminishing the potential across \lao.~\cite{Pickett2010review}  Thus within this picture, in the presence of an electron gas no field would be expected across the \lao.  However, if all the charge carriers do not lie precisely at the interface or have an extrinsic (oxygen vacancies~\cite{siemons2007} or cation doping~\cite{willmott2007}) origin, the \lao ~potential will not be fully screened and can thus be probed.~\cite{Li2009}  The precise band alignment between the \lao ~and \sto ~will also determine the strength of the residual fields in the \lao.~\cite{gu2009}  Addressing this issue and determining the existence of an uncompensated built-in potential in \lao ~(or polar ionic insulators in general) is central to understanding surface reconstruction phenomenon in thin films.~\cite{noguera2008a,goniakowski2007}  %Furthermore, band bending caused by a metal film deposited on the \lao ~may partly deplete the electron gas, giving rise to a measureable built-in potential.   When at least four unit cells of \lao ~are deposited on TiO$_2$ terminated \sto, an electron gas forms near the interface in \sto ~while the SrO terminated \sto ~remains insulating for any number of \lao ~overlayers.~ are spread across a finite depth or are buried within the \sto, the \lao ~potential will not be fully screened and can thus be probed.  The same reasoning holds if the charge carriers 

We probe the potential landscape across the \lao ~and the interface region in \sto ~by employing a typical metal-insulator-metal capacitor geometry such that the \lao ~thin films form the dielectric layer sandwiched between evaporated metallic electrodes and the electron gas, as depicted in Fig.\ 1a and described in Methods.  A bias voltage is applied to the top metallic electrode while the electron gas is held at ground.

A typical current-voltage ($JV$) curve measured between a Pt top electrode and the electron gas for an \lao ~thickness, $d_\mathrm{LAO}$, of 20 u.c. is shown in Fig.\ 1b and overlayed with theoretical curves, as labeled.  Similar rectifying $JV$ curves were measured for nine different LaAlO$_3$ films with $d_\mathrm{LAO}$ ranging from 5 to 40 u.c.  Junctions thinner than $d_\mathrm{LAO}$ = 20 u.c. are well described by direct tunneling between the metal electrode and electron gas for both positive ($+V$) and negative ($-V$) applied biases, as described in detail in Supplementary Section I (SSI).  Figure 1b shows excellent agreement to the theoretically calculated curve (black) for direct tunneling for $+V$.  For samples with $d_\mathrm{LAO}$ $\geq$ 20 u.c. however, an additional contribution from Zener tunneling for negative applied bias must be included, as shown in Fib. 1b.  %As described in detail in Supplementary Section I (SSI), the metal/\lao/\sto ~junctions can be modeled as standard metal-insulator-metal (MIM) tunnel junctions with different metal work functions.  

The inset in Fig.\ 1c shows a sampling of the $JV$ curves for various $d_\mathrm{LAO}$ with emphasis on the negative bias region at 100~K.  It is clear that for a given voltage the current does not decrease monotonically with $d_\mathrm{LAO}$.  The main panel of Fig.\ 1c shows $J$ as a function of $d_\mathrm{LAO}$ (open circles) at $V$ = -0.1~V for all samples, revealing a clear peak at 20~u.c.\ (7.56~nm).  The same trend is observed at all temperatures at negative bias.  The dark blue curve in Fig.\ 1c shows calculated values for $J$ vs. $d_{\mathrm{LAO}}$ within the direct tunneling model (see SS1), which also agrees well with the data for thicknesses below 20~u.c..
%
%To understand the observed thickness dependence, we recall that in the presence of a large enough electric field, Zener breakdown occurs across insulators and electrons tunnel from the valence to conduction band.  For materials with a built-in polar field, the internal potential grows with increasing thickness until a critical thickness is reached such that the built in potential is equal to the band gap.  Thus at the critical thickness, the valence and conduction bands align, giving rise to a significant increase in tunneling current.  
%
%The red lines in Fig.\ 1c are theoretically calculated current densities as a function of $d_{\mathrm{LAO}}$ for direct and Zener tunneling across polar \lao, as labeled and described below.  

The sudden increase of tunneling current at $d_\mathrm{LAO}$ = 20~u.c.\ seen in  Fig.\ 1c can be understood by considering the polar nature of \lao.  We recall that in the presence of a large enough electric field, Zener breakdown occurs in insulators and electrons tunnel from the valence to conduction band.~\cite{zener1934}  For materials with a built-in polar field, the internal potential grows with increasing thickness until it equals the bandgap at a critical thickness, $d^{cr}$, giving rise to a significant increase in tunneling current.~\cite{simon2009}  For \lao ~thin films deposited on TiO$_2$ terminated \sto, in the absence of complete screening an ionic built-in potential normal to the interface, pointing away from the \sto ~is expected.~\cite{nakagawa2006,Pickett2010review}  Neglecting effects of the Pt electrode, Figs.\ 2a and 2b illustrate zero bias band diagrams (black outline) where the band bending in LaAlO$_3$ reflects the unscreened potential.~\cite{nakagawa2006,gu2009,noguera2008a}   In green, we also sketch the cases for positive ($+V$, Fig.\ 2a) and negative ($-V$, Fig.\ 2b) applied bias.  When a metal film is brought into contact with \lao ~the built-in potential may be further modified depending on the precise band alignments which are difficult to accurately predict in the presence of electronic reconstruction.~\cite{gu2009}%additional bending in the \lao ~is expected to take place depending on the metal work function, which may modify the total built-in potential.~\cite{gu2009}  % as illustrated in Fig.\ 2c.~\cite{gu2009}  Larger bending and thus a larger reduction of the built-in potential is expected for larger work function metals.

The band diagram shown in Fig.\ 2a reflects the tunneling of electrons from the \sto ~to the Pt electrode (not shown for simplicity).  At large enough bias, electrons tunnel to the empty LaAlO$_3$ conduction band (Fowler-Nordheim tunneling~\cite{Fowler1928}) for applied bias V, when $qV \geq qV_{bi}  + {\Delta}E_{C}$, where $V_{bi}$ is the net built-in potential across LaAlO$_3$ including contributions from both the ionic built-in potential \textit{and} contributions from the metal work function. ${\Delta}E_{C}$ is the conduction band offset between the \sto ~potential well and \lao.  

%%The band diagram shown in Figure 2a reflects the direct tunneling of electrons from the \sto ~to the Pt electrode (not shown for simplicity).  At large enough bias, electrons tunnel to the empty LaAlO$_3$ conduction band (Fowler-Nordheim tunneling~\cite{Fowler1928}) when
%\begin{equation}
%    \label{eq4}
%    qV \geq qV_{bi}  + {\Delta}E_{C},
%\end{equation}
%where $V_{bi}$ is the net built-in potential across LaAlO$_3$ including contributions from both the ionic built-in potential \textit{and} band bending resulting from the metal work function; ${\Delta}E_{C}$ is the conduction band offset between the \sto ~potential well and \lao.  Figure 1b shows excellent agreement to the theoretically calculated curve (black) for direct tunneling, with the onset of Fowler-Nordheim tunneling for high enough positive bias as described in SS1.  The dark blue curve in Fig.\ 1c shows calculated values for $J$ vs. $d_{\mathrm{LAO}}$ within the direct tunneling model (see SS1), which also agrees well with the data.

Figure 2b shows electrons tunneling from the metal into the \sto ~well, as determined from the data shown in Fig.\ 1b.  Above a critical \lao ~thickness, $d^{cr}_\mathrm{LAO}$, we also expect electron tunneling across the \lao ~bandgap, via the valence states for applied bias $V$ when:
\begin{equation}
    \label{eq1}
    qV \geq E_{g(LAO)}-qV_{bi}.
\end{equation}
Here $E_{g(LAO)}$ is the bandgap of \lao.  Using the diagram in Fig.\ 2b, the current density of electrons tunneling across the LaAlO$_3$ bandgap was calculated within the WKB approximation~\cite{zener1934,simon2009} as shown in SSI, using
\begin{equation}
    \label{eq5}
\Phi(x)= E_{g(LAO)}-qE_{bi}d_{\mathrm{LAO}}-qV+\epsilon 
\end{equation}as the potential barrier height, where $\epsilon$ is the total kinetic energy of the tunneling electrons, $E_{bi}$ the effective built-in electric field in \lao, and $V$ = -0.1~V.  We obtain the light blue curve shown in Fig.\ 1c, which is in excellent agreement with the data using the experimentally measured $E_{g(LAO)}$ = 6.5~eV~\cite{mi2007} for \lao ~grown on \sto.   

A distinguishing feature of polar tunnel barriers is the low tunnel probability just below $d^{cr}$.~\cite{simon2009}  Our measurements (Fig.\ 2c) indicate that $d^{cr}_{\mathrm{LAO}}$ lies somewhere between 17~u.c.\ and 20~u.c., or 18.5 $\pm$ 1.5~u.c.\ when using Pt electrodes.  Thus, using $d^{cr}_{\mathrm{LAO}}$ = 18.5, $a$ = 0.378~nm~\citep{maurice2006} for the lattice constant of \lao,  and $E_{bi}$$d^{cr}_{\mathrm{LAO}}a$ = $qV_{bi}$ = $E_{g(LAO)}$ = 6.5~eV (from Eq.~\ref{eq1} with $qV$ = 0), we obtain $E_{bi}$ = 93.0~meV/\AA ~as an estimate for the built-in electric field across LaAlO$_3$.  $E_{bi}$ reduces to 50.3~meV if tunneling from the \lao ~valence to the \sto ~conduction band is assumed instead, as discussed in SS1 as an alternate scenario.  Alternatively $E_{bi}$ = 80.1 meV/\AA ~if the bulk value, $E_{g(LAO)}$ = 5.6~eV is used instead.  A thorough metal work-function dependence is needed to accurately determine the ionic built-in potential contribution to $V_{bi}$.  We note that band bending in the \lao ~from the metal alone cannot account for the observed thickness dependence.  Although,we expect band bending from the Pt electrode to modify the measured $V_{bi}$, it is not clear what fraction of the ionic potential is partially screened by the electron gas or partially compensated by covalency within the \lao.~\cite{Pickett2010review,noguera2008a}   %Band bending effects can convolute matters further. These numbers are close to the value of $V_{bi}$ obtained in a recent photoemission study and are much smaller than the values expected using a 4~u.c.\ critical thickness for \lao ~(giving $E_{bi}$ = 235~meV/\AA ~using Supplemental Eqn.\ 8), hence the presence of a polar field in \lao ~was ruled out in that study.~\cite{segal2009}   18.5uc = 69.93 A

It is interesting to note that $d^{cr}_{\mathrm{LAO}}$ $\approx$ 18.5~u.c.\ for tunnel-coupling of the two \lao ~surfaces (in the presence of a Pt electrode) is close to the value of 15 u.c.\ (using no metal electrode) above which a drop in the in-plane mobility~\cite{bell222111} and change in the in-plane magnetoresistance~\cite{bell222111,brinkman2007} have previously been observed.  Any hole gas present in the \lao ~valence band for $d_{\mathrm{LAO}} \geq d^{cr}_{\mathrm{LAO}}$ will contribute to the measured hall resistance and calculated mobility (since contacts made to the \sto ~gas are connected through the \lao ~film), while the cross-over from positive to negative magnetoresistance  for $d_{\mathrm{LAO}} \geq d^{cr}_{\mathrm{LAO}}$ may reflect bilayer effects similar to those observed in tunnel-coupled quantum wells.~\cite{ihn1996}%bears intriguing resemblance to magnetic-field induced resistance resonance in tunnel-coupled quantum wells with different carrier mobilities.~\cite{ihn1996}  %*** while the magetoresistance will surely reflect the onset of a parallel conduction channel.

Our data clearly reveal that the electron gas does not fully screen the ionic built-in potential across \lao.  This suggests a couple of possibilities:  (i) the charge carriers do not all lie exactly at the interface and are buried one or a few monolayers within the \sto, or (ii) the metal electrodes deposited on the \lao ~modify the band alignments and thus the carrier density of the electron gas.  The latter effect may explain the failure to detect a built-in potential using photoemission experiments conducted in the absence of metal electrodes.~\cite{segal2009,gu2009}  Although the present work cannot distinguish between the above scenarios, to the best of our knowledge uncompensated built-in electric fields across ionic insulators with polar terminations have not previously been detected.~\cite{Goniakowski2008,gu2009}  The resulting potential is analogous to the built-in potential that arises in semiconductors due to free charge separation and the switchable built-in potential found in ferroelectrics, generally resulting from covalently bonded dipoles.  The ionic built-in potential is most similar to the potential found in piezoelectric materials which also results from covalent bonds and is tuneable.  However, unlike the piezoelectric/ferroelectric potential, the ionic built-in potential is unique in that it results from permanent dipoles.% which cannot be tuned with applied electric fields.

%much smaller than the values expected in the ionic limit
%MOVE to supplementary information
%In SSI, we consider the alternate scenario of tunneling from the metal electrode into the \lao ~conduction band (Fowler-Nordheim tunneling) and why Zener tunneling is the most likely scenario.    In either case, the thickness dependent measurements reveal the presence of an intrinsic polar field across \lao.  
%XXXBegin ConstructionXXX
%
%Next, we explore the effects of the \sto ~dielectric permittivity on the tunneling current.  significant temperature dependence observed for positive biases (direct tunneling) as shown in Fig. 2a for a 20~u.c. sample.  
%
In addition to revealing a built-in potential across \lao, the tunneling data also reveal information about the band profile in the \sto ~potential well region.  The \sto ~interface region can be probed by tuning both the charge density and the \sto ~dielectric permittivity, $\chi$$_{STO}$, with temperature and applied electric fields.~\cite{Copie2009,Bell2009a,siemons2007}  If the \lao/\sto ~heterostructure is compared to the inversion layer in a lightly doped metal-oxide-semiconductor (MOS) capacitor, we can expect changes in the charge density of the inversion layer as a function of temperature and gate voltage.~\cite{sze2006}  Changes in either of these quantities will be reflected in the barrier height, $\Phi$ for electrons tunneling $from$ the \sto ~potential well. %Changes in the charge density will affect the interface band bending in \sto, and thus also $\Delta$$E_C$ as shown in the inset of Fig.\ 3a.  Further, changes in $\chi$$_{STO}$ near the interface induced by applying a field across the \sto ~will affect $\Delta$$E_C$ by changing the confining potential.  Changes in $\Delta$$E_C$ will be reflected in the barrier height, $\Phi$(x) for electrons tunneling $from$ the \sto ~potential well. 

%For a better understanding of the \sto ~interface region, it is necessary to account for the temperature and field dependence of the \sto ~dielectric permittivity, $\chi$$_{STO}$.  For unstrained, bulk \sto ~crystals, $\chi$$_{STO}$ grows by a factor of 10$^2$ as the temperature is decreased from 300~K to 10~K.~\cite{muller1979}  Additionally, $\chi$$_{STO}$ has a complex dependence on electric fields, $E$, namely, $\chi$$_{STO}$ = $\frac{C1}{1-E/C2}$ where $C1$ and $C2$ are constants.~\cite{Copie2009}  Recent works focused on the influence of $\chi$$_{STO}$ on the \lao/\sto ~interface suggest that the electron gas reduces (grows) in volume as $\chi$$_{STO}$ increases (decreases), while the change in charge density is relatively small.~\cite{Copie2009,Bell2009a,siemons2007}  The same studies when combined also suggest that a net change in $\chi$$_{STO}$ at the interface reflects the dependence on temperature, fields applied across \sto ~and the interface confining potential.~\cite{Copie2009}   Thus, we expect that field or temperature dependent changes in both the electron gas volume and charge density will affect the interface band bending and thus $\Delta$$E_C$ (Inset, Fig.\ 3a). This in turn will be reflected in the barrier height, $\Phi$(x), for electrons tunneling $from$ the \sto ~potential well.  

In order to probe changes in the barrier height, we tune $\chi$$_{STO}$ and the charge density using both temperature and an applied field across \sto, and measure the resulting tunneling curves as shown in Figs.\ 3a and 3c.  In agreement with the discussion above, we find that in the $-V$ region where electrons are tunneling from the metal electrode via the \lao ~valence band $to$ the \sto ~potential well, $JV$ undergoes little change while significant changes are observed for electrons tunneling from the electron gas in \sto ~for $+V$.

%In order to probe changes in $\Delta$$E_C$, we tune $\chi$$_{STO}$ and the charge density using both temperature and an applied field across \sto, and measure the resulting tunneling curves as shown in Figs.\ 3a and 3c.  In agreement with the discussion above, we find that in the $-V$ region where electrons are tunneling from the \lao ~valence band and the metal electrode $to$ the \sto ~potential well, $JV$ undergoes little change while significant changes are observed for electrons tunneling from the electron gas in \sto ~for $+V$.

Both the backgate and temperature dependence data are qualitatively in agreement with the simple barrier height change arguments presented above.  Fig.\ 3a shows $JV$ curves for the 20~u.c.\ sample at several temperatures, while Fig.\ 3b shows changes in $\Phi$ as a function of temperature as obtained by fitting to the direct tunneling model (SS1).  As fewer states are occupied in the interface region with decreasing temperature, the barrier height of electrons tunneling from the \sto ~increases.  Similarly, Fig.\ 3c shows $JV$ for the 30~u.c.\ sample at 50~K for several back gate fields ($E_{bg}$) applied with respect to the electron gas, as schematically depicted in the inset.  Fig.\ 3d shows the corresponding changes in $\Phi$.  This is in agreement with previous observations~\cite{Bell2009a} where negative $E_{bg}$ compresses the electron gas while also reducing the charge density.  We do not observe significant changes in the $\Phi$ for positive $E_{bg}$.  This is expected as the electron gas has a higher charge density and occupies a larger volume.~\cite{Bell2009a}  Interestingly, the change in $\Phi$ begins to level off below about 50~K in Fig.\ 3b.  Since \sto ~is an incipient ferroelectric and is sensitive to the slightest strain perturbation, it is possible that we are observing the effects of an interface dipole emerging at low temperatures.~\cite{muller1979}  A change in the dipole strength at the interface with temperature and an applied back gate will modify $\Delta$$E_C$ which can also change $\Phi$.

Capacitance measurements are an excellent tool for probing this latter notion.  Fig.\ 4a shows $CV$ measured between the metal electrode and electron gas at 10~K for a 30~u.c.\ sample.  (Supplementary Section II, or SSII, provides details about the complex impedance analysis).  The onset of Zener tunneling at negative biases coincides with a sharp drop in $C$, which is accompanied by a sharp drop in the complex phase angle:  Figure 4b shows the phase angle, $\delta$, of the complex impedance as a function of $V$ at several temperatures.  $\delta$ is calculated to be exactly 90$^{\circ}$ for a perfect capacitor, becoming less than 90$^{\circ}$ as the dielectric becomes more conducting (SSII).  As seen here, the drop off to zero in $\delta$($V$) coincides with the sharp drop in $CV$ and the onset of Zener tunneling.  During Zener tunneling charge carriers are introduced into the \lao ~conduction band and \lao ~is no longer a good dielectric, behaving more like a resistor.  When $\delta$ = 90$^{\circ}$ however, the system behaves like a metal-insulator-metal tunnel junction.  For samples thinner than 20~u.c., while we do observe a typical MOS capacitor depletion capacitance at high enough negative bias, no corresponding drop in the complex phase angle is observed (SSII).  %%The same behavior is observed for all samples with thicknesses $\geq$ 20~u.c..  Qualitatively, the $CV$ show that the metal/\lao/\sto ~structure can be thought of as a reverse polarity or $backward$ MOS capacitor, akin to backward pn junctions which exhibit Zener tunneling.~\cite{sze2006}The nearly constant $CV$ when $\delta$ = 90$^{\circ}$ confirm the robust dielectric properties of  \lao ~and rule out semiconductor effects such as the presence of mobile charge or excessive defect doping in \lao. 
%
%When $\delta$ = 90$^{\circ}$, $C_P$ = $C_S$., the sample behaves like an ideal metal-insulator-metal capacitor and is model independent.  When $\delta$(V) in Fig.\ 3d deviates from 90$^{\circ}$,  $C_P$ and $C_S$ diverge indicating increased conduction across \lao.  The onset of Zener tunneling corresponds to the drop in $\delta$(V).  The $CV$ measurements when combined with $JV$ confirm the robust dielectric properties of  \lao ~and rule out semiconductor effects such as the presence of mobile charge or excessive defect doping in \lao. 

Figures 4b to 4d provide a closer examination of the hysteretic behavior that appears below 100~K in both the $CV$ and $\delta$($V$).~\cite{Bell2009a}  We find that the memory window defined by the hysteresis remains constant at $\Delta$$V$ = 0.16~V for frequencies below the $RC$ roll off when $V_{max}$ = 0.8~V  (Fig.\ 4b), and increases with decreasing temperature (Fig.\ 4c).  The memory window also increases with increasing maximum applied voltage ($V_{max}$ in Fig.\ 4d).  The lack of a frequency dependence and decrease in hysteresis with increasing temperature rule out the role of interface traps which are more active at higher temperatures and have a characteristic time dependence associated with charging/discharging.~\cite{sze2006}  

The data however compare well to the response measured for metal-ferroelectric-insulator-metal or semiconductor (MFIM or MFIS) junctions.~\cite{liu2002,miller1992}  Using the change in charge density for the high and low $C$ values shown in Fig.\ 4a, we obtain 18 nC/cm$^{2}$ as an estimate of the remnant polarization for a maximum applied voltage of 0.8~V.  Since ferroelectric behavior in LaAlO$_3$ has never been observed or expected, any dipole switching can be attributed to a thin layer in SrTiO$_3$ near the interface.~\cite{bickel2009}  This also suggests that the electron gas lies not exactly at the interface, but burried one or a few monolayers within the \sto.  It can thus be argued that the large influx of charge carriers into the potential well at the onset of Zener tunneling changes the local electric field and modifies the dipole strength.  %Hence, the onset of Zener tunneling is associated with hysteretic $CV$.

These observations support recent experimental evidence of an induced polarization with bound charges in addition to free charge near the interface in \sto.~\cite{ogawa2009,salluzzo2009}  Furthermore, there is evidence from both X-Ray~\cite{vonk2007} and TEM~\cite{maurice2006} measurements that when LaAlO$_3$ with a 3\% lattice mismatch is deposited on the (001) SrTiO$_3$ substrates, distortions in the TiO$_6$ octahedra occur at the interface.  It is also well known that biaxial strain as well as a strain gradient can induce ferroelectric~\cite{haeni2004} and flexoelectric~\cite{zubko2007} polarizations in \sto.  The resulting bound charges would give rise to a finite electric field and confining potential at the interface.~\cite{pentcheva2008,yoshimatsu2008,glinchuk2004}  This is analogous to the mechanism that gives rise to a confined electron gas in wide band gap GaN based heterostructures.~\cite{ambacher2000}  %Curiously, a recent calculation that also accounts for the complex electric field dependence of the SrTiO$_3$ bulk permittivity shows that screening charges accumulate near the SrTiO$_3$ surface when a finite electric field is present at the interface that decays to zero in the bulk substrate.~\cite{Copie2009}

Thus, in addition to polar \lao, the weak polarity of the \sto ~appears to play an important role in determining the electrostatic boundary conditions at the \lao/\sto ~interface.  Theoretically, it was shown that depending on the termination layer, TiO$_2$ or SrO, (001) bulk \sto, which is iono-covalent and weakly polar, is expected to undergo weak electronic or structural reconstructions near the surface to avoid a diverging potential.~\cite{Goniakowski1996}  Very recently, it was shown that when polar \lao ~overlayers are added to polar \sto, the accumulation of screening charges is one means for ensuring conservation of the electric displacement vector across the \lao/\sto ~interface.~\cite{massimiliano2009}  %In this respect, it is also possible SrTiO$_3$ responds differently to strain induced fields with a TiO$_2$ termination than with the SrO termination.  This could potentially help explain the lack of mobile charge carriers at the SrO terminated interface.   %INSERT SOMETHING ABOUT THE STO SUBSTRATE ACCOMODATING STRAIN RATHER THAN STRAINING THE THIN FILM.  ***Talk about how this is something that is worth expoloring more seriously in trying to understand this electron gas.***For instance, by stipulating a continuous electric displacement at the surface, self-consistent solutions to the Schrodinger-Poisson equation show that depending on the termination layer, TiO$_2$ or SrO, (001) bulk \sto, which is iono-covalent and weakly polar, is expected to undergo weak electronic or structural reconstructions near the surface to avoid a diverging potential

To summarize, using capacitance and electron tunneling measurements we have provided experimental evidence of a finite built-in electric field across polar \lao.  We have also shown evidence of a finite polarization and bound charge within \sto, highlighting a unique property of this oxide substrate.  These two results provide key insights to understand and design novel interfaces in a controlled manner.  Furthermore, built-in fields in large band gap polar insulators could potentially be exploited for applications such as reducing leakage current in transistors,~\cite{cao2010} achieving high power MOS capacitors and electron hole separation for enhancing built-in fields in pn junction diodes,~\cite{simon2009} etc.  Essentially, a built-in field in singly terminated polar ionic insulators results from permanent dipoles while for non-polar doped semiconductors physical separation of positive and negative free charge carriers gives the same effect.  As semiconductor device concepts achieve maturity, much of the same phenomenology can be translated to the complex oxides which offer many more degrees of freedom for exploration with applications perhaps not yet conceived.  %Our data also unveil an avenue for designing \textit{active} electronic devices utilizing insulating oxides which are generally considered \textit{passive} components. 

\section*{Methods}
The \lao/\sto ~samples were fabricated by epitaxially depositing \lao ~films on TiO$_2$ terminated (001) \sto ~substrates by pulsed laser deposition using a KrF laser. Before growth, the substrates were preannealed at 1223~K for 30~min in an oxygen environment of 0.67~mPa. Following this anneal the growth was performed at 1073~K in an oxygen pressure of 1.33~mPa, at a repetition rate of 2~Hz. The total laser energy was 20~mJ, and the laser was imaged to a rectangular spot of area approximately 2.3 x 1.3~mm$^2$ on the single crystal \lao ~target using an afocal zoom stage. After growth, the samples were cooled to room temperature in an O$_2$ pressure of 4 $\times$ 10$^4$~Pa, with a one hour pause at 873~K.~\cite{Thiel2006,Bell2009a} Nine samples with \lao ~thicknesses of 5, 7, 10, 13, 15, 17, 20, 30 and 40 u.c.\ were deposited in this way.  

Several circular metallic electrodes with diameters ranging from 0.3 to 0.7~mm were thermally evaporated on each of the \lao ~films using a shadow mask.  Au was used for the 30~u.c.\ sample and Pt for all other samples.  The backgate was thermally evaporated onto the 30~u.c.\ sample.  Gold wire (0.0025" gauge) was manually bonded to the electrodes using silver epoxy and an Al wirebonder was used to bond to the electron gas.  The samples were cooled in Quantum Design PPMS and MPMS systems.  Tunneling measurements were conducted in Kashiwa, Gainesville and Berkeley, using various electrometers and source-measure units, all giving the same results.  Capacitance measurements were also performed in all three locations using an HP 4284 LCR meter.  For the back gate measurements, electric fields were applied in the following succession: 0~kV/cm $\rightarrow$ negative fields $\rightarrow$ positive fields.

%%%%%%%%%%%%%%%
%********************BIBLIOGRAPHY************

%%%%%%%%%%%%%%%%%
%%%%%%%%%%%

\section*{Acknowledgments}
We thank M. Gajek and J. H. Bardarson at UC Berkeley for discussions.  GSB acknowledges support from the Japan Society for Promotion of Science (Award No. SP08057) and the U.S. National Science Foundation (Award No. OISE0812816) under the 2008 EAPSI fellowship program.  The work at Berkeley (RR) was supported by the US Department of Energy under contract No. DE-AC02-05CH1123.  The work at Florida
(AFH) was supported by the US National Science Foundation under Grant No. 0404962.  WS acknowledges support from the Dutch Organization for Scientific Research (NWO-Rubicon Grant).
\section*{Author Contributions}
CB deposited the \lao ~films. GSB prepared and measured the tunnel junctions with CB, and modeled/analyzed the data with JR and WS.  SS simulated the $JV$ curves within the NEGF approach.  The manuscript was prepared by GSB with assistance from CB, JR, WS and YH.  HYH, AFH and GSB contributed to conceptualizing the experiment.  HYH provided insights and expertise on the \lao/\sto ~interface, RR on ferroelectricity and AFH on interpreting complex impedance.
\section*{Competing Financial Interests}
The authors declare no competing financial interests.
\section*{Figure Legends}

\subsection*{Figure 1}
\textbf{Built-in polarization across \lao/\sto ~tunnel junction diodes \textbar ~a,} Tunnel junctions are formed between thermally evaporated circular metallic electrodes on \lao ~and the electron gas, as described in Methods.  \textbf{b,} $JV$ curve measured at 10~K for a 20~u.c.\ sample compared to curves calculated within the direct and Zener tunneling models as labeled, and described in SSI.  \textbf{c,}  $J$ vs.\ \lao ~thickness, $d_{\rm{LAO}}$, for all samples at $V$ = -0.1~V showing a clear peak at 20~u.c. (7.56~nm).  Calculated curves for direct tunneling (dark blue) and Zener tunneling (light blue) are labeled and shown.  The inset provides a closer look at the negative bias region of $JV$ for several thicknesses showing that $J$ does not scale monotonically with thickness for a given $V$ (see dotted line at $V$ = -0.1~V). 
\subsection*{Figure 2}
\textbf{Thickness dependent built-in potential and inter-band tunneling across polar \lao ~\textbar ~a,}  A schematic band diagram of the \lao/\sto ~interface at zero bias (black outline) showing the onset of Fowler-Nordheim tunneling for positive applied bias ($+V$, green outline) and, \textbf{b,}  the onset of Zener tunneling at negative applied bias ($-V$, green outline) at a critical thickness across polar \lao ~with a built-in electric field. % \textbf{c,}  Additional band bending across \lao ~due to the metal electrode.  The larger the metal work funciton, $\Phi_m$, the larger the bending down of the \lao ~bands on the metal side and the smaller the measured $V_{bi}$.
\subsection*{Figure 3}
\textbf{Tuning the tunneling current across \lao ~by tuning the \sto ~permittivity and charge density \textbar ~a,} $JV$ curves for a 20~u.c.\ \lao ~sample are shown for several temperatures with theoretical fits to the direct tunneling model (black).  The inset schematically depicts how changes in the Fermi level with temperature or applied back-gate fields ($E_{F1}$ at 300~K and $E_{F2}$ at 10~K) affect the charge density in the \sto ~potential well region and hence the barrier height $\Phi$.  {\bf b,}  Barrier heights extracted from fits to the direct tunneling model for each of the $JV$ curves shown in {\bf a}.  {\bf c,}  $JV$ for a 30~u.c.\ \lao ~at 50~K for several different positive and negative applied \sto ~back-gate fields, $E_{bg}$.  The effect is much more pronounced for negative rather than positive biases. {\bf d,} Each of the $JV$ curves shown in {\bf c} was also fit to the direct tunneling model.  The barrier height increases linearly with increasingly negative backgate fields. %indicating again that as electrons become confined closer to the interface with  increasingly negative fields~\cite{Bell2009a} the band bending increases, which in turn increases $\Delta$$E_C$ and $\Phi$(x). The barrier height increases linearly with decreasing temperature, indicating that as electrons become confined closer to the interface with decreasing temperature (shown schematically in the inset of {\bf a}) the band bending increases, which in turn increases $\Delta$$E_C$ and thus $\Phi$(x). 
\subsection*{Figure 4}
\textbf{Capacitance measurements agree with $JV$ while also revealing an induced dipole in \sto ~\textbar ~a,} $CV$ curves for a 30~u.c.\ sample measured at 10~K and 10~kHz is qualitatively similar to a metal-insulator-semiconductor (MIS) capacitor curve with a ferroelectric contribution, or a metal-insulator-ferroelectric-semiconductor (MIFS) capacitor.  The drop in $CV$ occurs during the Zener tunneling regime when carriers are introduced into the \lao ~conduction band, making it a leaky dielectric. {\bf b,}  The phase angle of the measured complex impedance as shown at several temperatures.  A hysteresis appears near 100~K and increases with decreasing temperature. {\bf c,} The hysteresis is frequency independent for frequencies below the $RC$ roll-off limit, and {\bf d,} the hysteresis window, $\Delta$V, increases with the maximum applied voltage, $V_{max}$, for a given sweep.  {\bf b} to {\bf d} together qualitatively provide strong indications of dipole switching at the interface in \sto.

%\begin{references2}

\end{document}

\bibliographystyle{naturemag}

\newcommand{\lao}{{LaAlO$_3$}}
\newcommand{\sto}{{SrTiO$_3$}}
\title{Supplementary Information: Built-in and induced polarization across \lao/\sto ~heterojunctions}

\author{}

\maketitle

\textbf{Section I.  Metal/L\lowercase{a}A\lowercase{l}O$_3$/S\lowercase{r}T\lowercase{i}O$_3$ tunnel junctions}

In order to identify the mechanisms governing the observed rectifying behavior in $JV$ (see Fig.\ 1c of the main text), we have fit the data to standard semiconductor heterostructure and tunnel junction models, namely:  Thermionic emission, Poole Frenkel conduction, space charge and diffusion limited currents, direct tunneling, Fowler-Nordheim tunneling, Zener tunneling and trap assisted tunneling.~\cite{sze2006}  We have found that our data are best described by the tunneling models, each dominating within a certain voltage range depending on the built-in potential (shown schematically in Fig.\ 2a and 2b of the main text).  The total current density measured across the metal/\lao/\sto ~tunnel junctions is given by 
\begin{equation}
    \label{eq2}
    J=J_{FN}+J_{DT}+J_Z,
\end{equation}
where $J_Z$ is the interband tunneling current between the \lao ~valence and conduction bands, $J_{FN}$ is Fowler-Nordheim tunneling current across a triangular barrier with electrons tunneling from the metal or electron gas into the empty \lao ~conduction band and $J_{DT}$ is the direct tunneling current across a trapezoidal barrier between the metal and electron gas.  To determine the effective mass, the $J_Z$ curves (Fig.\ 1c main text) were first calculated using the measured \lao ~thin film band gap and varying the effective mass.  The effective mass that gave best agreement with the data was used to determine the barrier heights for $J_{DT}$.  

We also simulate $JV$ curves using the non-equillibrium Green's function approach for the band diagram shown in Figs.\ 2a and 2b of the main text.  We obtain excellent qualitative agreement with the data as shown in the section labeled ``Two band model simulations" below.

\begin{figure}
\includegraphics[width=\columnwidth]{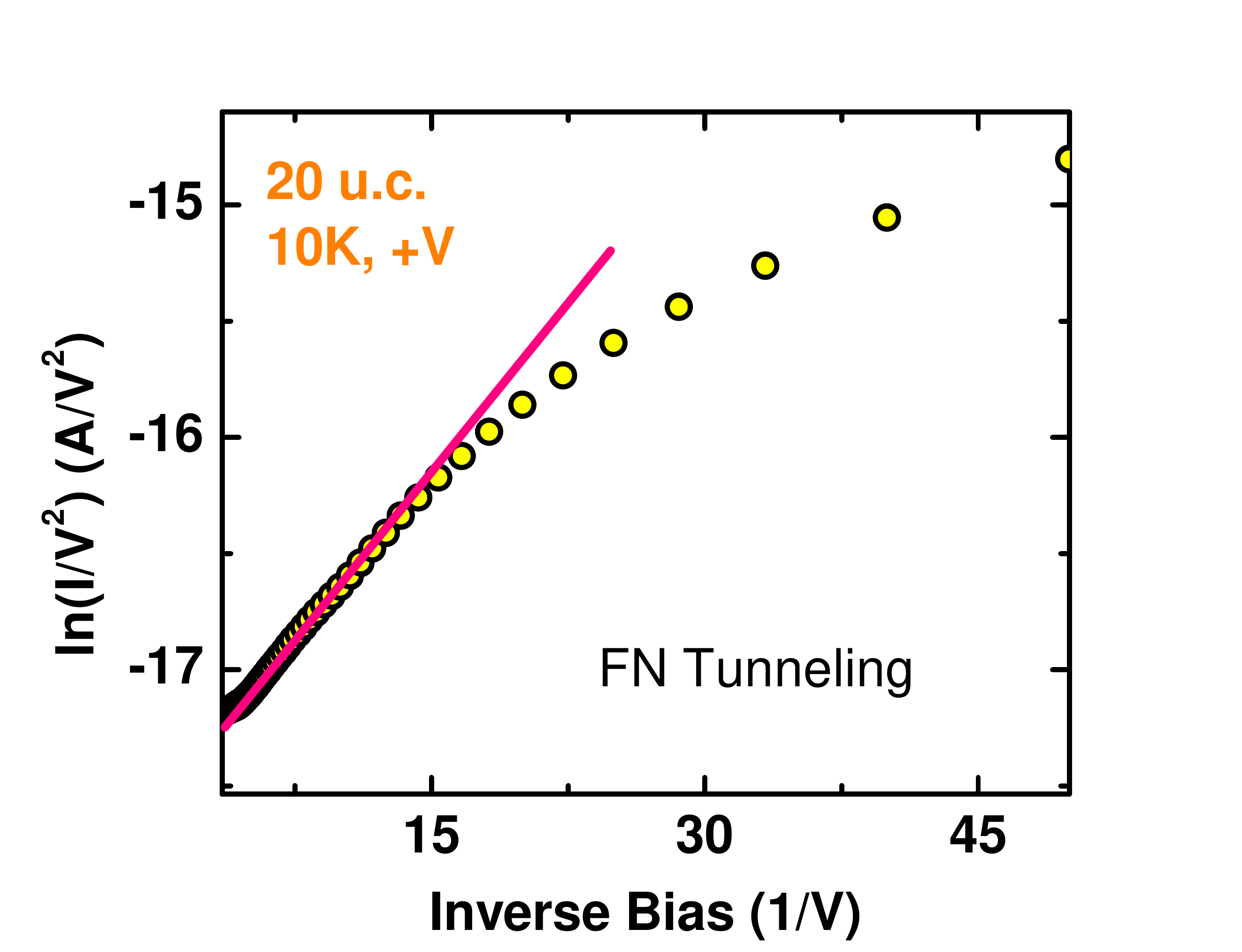}
\caption{For the higher positive applied voltages ($+V$), Fowler-Nordheim tunneling describes the data well for the 20~u.c. sample shown.}
\label{f2}
\end{figure}
%We find that when the effective mass is set as a fit parameter, it converges to  $m^*$ = (0.153 +/- 0.010)$m_\circ$ within all the different tunneling regimes.  Thus $m^*$ was fixed at 0.153$m_\circ$ for all tunneling fits and the barrier height was set as the fit parameter.  

\textbf{Direct~tunneling}:  For direct tunneling we use the Simmons model~\cite{simmons1963} which uses an average barrier height since the metal electrode and electron gas are expected to have different work functions.~\cite{wolf1985}

    \begin{align}
    \label{eq2}
    J_{DT} &=   \frac{e}{2\pi h(x\beta)^2}\times \notag \\ 
    &\left\{(\overline{\phi}-\frac{eV}{2})   \exp[   \frac{-4\pi \beta d_{\rm{LAO}}(2m^*)^{1/2}}{h}  (\overline{\phi}-\frac{eV}{2})^{1/2}] \right.  \\
   & \left. -(\overline\phi+\frac{eV}{2})
    \exp[\frac{-4\pi \beta d_{\rm{LAO}}(2m^*)^{1/2}}{h}   (\overline{\phi}+\frac{eV}{2})^{1/2}]\right\}. \notag
   \end{align}
 
Here $\overline{\phi}$ is the average barrier height, $d_{\rm{LAO}}$ is the \lao ~thickness and $\beta$ is an ideality factor that was set to 1 for all our calculations.  In Fig.\ 1b of the main text, we have simulated the exact form using $m^*$ = 0.14$m_\circ$ as determined from the $J_Z$ calculation as discussed below.  For the thinnest samples, $\overline{\phi}$ $\approx$ 4~eV was obtained for positive applied voltage.  If the band gap of \lao ~in thin film form~\cite{mi2007} (6.5~eV) is used along with an \sto ~band gap of 3.2~eV, we obtain an \lao/\sto ~conduction band offset of 3.3~eV, suggesting that the calculated $\overline{\phi}$ is reasonable.  For the 20~u.c.\ sample on the other hand, significant barrier thinning and barrier height lowering due to the intrinsic \lao ~polarity, (shown in Fig.\ 1, main text) give $\overline{\phi}$  = 1.135~eV at 10~K, which increases by $\sim$ 0.01~eV for thicker samples (see Fig.\ 4, main text).  Interestingly, unlike typical metal-insulator-metal tunnel junctions, the barrier height for direct tunneling was found to be temperature dependent for positive biases (when tunneling from the \sto ~potential well, see Fig.\ 3b, main text), reflecting the similarities of this system to that of the inversion layer in a metal-oxide-semiconductor (MOS) capacitor which also exhibits temperature dependent tunneling currents.~\cite{sze2006}

%\begin{figure}
%\includegraphics[width=\columnwidth]{LAO-STO-figures-paper/Supp-Figs/Fig1.pdf}
%\caption{For the lower positive applied voltages ($+V$), direct tunneling fits the data well, as shown in Fig.\ 1 of the main text.  The same data are plotted here to show linear behavior during this regime when plotted as $I/V$ vs. $V^2$, an approximation of the Simmons' model.  The arrow indicates increasing voltages.}
%\label{f1}
%\end{figure}
%
As shown in Fig.\ 1b, only a portion of the $JV$ curve can be fit for negative applied bias ($-V$).  For the thinnest samples, we obtain good fits using Simmons' model for both $+V$ and $-V$ albeit with different barrier heights (reflecting the different work functions on either side of the \lao).  
%
%The dataset in Fig.\ 2b. of the main text was also fit to Eqn.~\ref{eq2} for the thinnest samples (red curve marked ``direct tunneling''), giving $\overline{\phi}$ = 4.0~eV.

\textbf{Fowler~Nordheim~tunneling}:  With a large enough field across \lao, the band bending is strong enough such that electrons can tunnel from the metal to the conduction band of \lao ~for $-V$ and, from the \sto ~potential well to the \lao ~conduction band for $+V$.  This type of tunneling, approximated with a triangular barrier across \lao ~is best described by the Fowler-Nordheim~\cite{Fowler1928} (FN) equation:

\begin{equation}
    \label{eq3}
    J_{FN}=\frac{e^2V}{16\pi^2\hbar \overline{\phi}d_{\rm{LAO}}} \exp[   \frac{-4\sqrt{2m^*}d_{\rm{LAO}}(q\overline{\phi})^{3/2}}{3\hbar eV}].
\end{equation}

The FN plot in Supplementary Fig.~\ref{f2} qualitatively shows that the high bias region for $+V$ is indeed described by FN tunneling for 20~u.c..  This can also be seen in Fig.\ 1c. of the main text, where the data points begin to deviate from the direct tunneling fit at high biases.  For most of our measurements, the high bias required to observe FN tunneling was not applied to avoid dielectric breakdown.

\textbf{Zener tunneling}:
%
%**********************
%MOVE TO SUPPLEMENTARY: To understand the observed thickness dependence of the tunneling current, we invoke an inter-band tunneling model.  In Zener diodes for instance, when applied electric fields exceed the band gap of the p-n junction, electrons tunnel from the valence band of the p-doped region into the conduction band of the n-doped region, depicted schematically in Fig.\ 2c.~\cite{sze2006}  The same mechanism gives rise to conduction (Zener breakdown) in insulators.~\cite{zener1934}  For materials with a built-in polar electric field on the other hand, regardless of the source of polarization (i.e. ionic, piezoelectric, etc.), the built-in potential grows with increasing thickness.~\cite{simon2009}  At a critical thickness the built-in potential is large enough such that the valence band and conduction band align without an external bias, giving rise to inter-band tunneling as shown in Fig.\ 2d.
%***********************

The thickness dependence of $J$ shown in Fig.\ 1c. of the main text can best be understood by considering interband tunneling as depicted in the band diagrams in Fig.\ 2a and 2b.  An alternate scenario is the tunneling of electrons from the \lao ~valence band to the \sto ~conduction band.  We note that both cases require \lao ~to be polar with a critical thickness that aligns either the \lao ~valence and conduction bands, or the \lao ~valence band with the \sto ~conduction band.  Thus the exact tunneling mechanism does not change our main result.  However, as shown below, our analysis indicates interband tunneling is the correct scenario.  

We obtain excellent fits to the data considering the first case of interband tunneling within the \lao.  Interband tunneling was first used to describe insulator breakdown by C. Zener~\cite{zener1934} and calculated within the WKB approximation:~\cite{zener1934,simon2009}
\begin{equation}
    \label{eq4}
    J_Z=\frac{em^*}{2{\pi}{\hbar}}\int_0^{\delta\epsilon} \! [f_v^{LAO}(\epsilon)-f_c^{LAO}(\epsilon)]{T}_{wkb} \, d\epsilon.
\end{equation}
Here, $J_Z$ is the interband tunneling current density. $f_v^{LAO}(\epsilon)$ and $f_c^{LAO}(\epsilon)$ are the Fermi-Dirac electron occupation functions for the valence and conduction bands in the \lao ~respectively.  This equation can easily be modified to include \lao ~to \sto ~tunneling by substituting $f_c^{LAO}$ with $f_c^{STO}$.  The tunneling probability ${T}_{wkb}$ in either case is given by,
\begin{equation}
    \label{eq5}
    {T}_{wkb}=e^{{-2\int_0^{d_{\mathrm{LAO}}} \! \sqrt{\frac{em^*}{{\hbar}^2}\Phi(x)} \, dx}}
\end{equation}
where for Zener tunneling in \lao, the potential barrier height, $\Phi(x)$, where $x=d_{\rm{LAO}}$ is given by
\begin{equation}
    \label{eq6}
\Phi(x)= E_{g(LAO)}-qE_{bi}x+\epsilon = E_{g(LAO)}(1-\frac{qx}{d^{cr}_{\rm{LAO}}})+\epsilon
\end{equation}
Here $\epsilon$ is the energy of the tunneling electrons, ${d^{cr}_{\rm{LAO}}}$ is the critical thickness from Fig.\ 2b., $E_{bi}$ = $E_{g(LAO)}/{d^{cr}_{\rm{LAO}}}$, the effective built in electric field in \lao, and $x$ the \lao ~thickness.  The function given in Eqn.~\ref{eq4} for small biases is approximately:~\cite{zener1934,simon2009}
\begin{equation}
    \label{eq7}
J_Z \approx \frac{e^3m^* {T}_{wkb}V^2}{4\pi^2\hbar}.
\end{equation}
Using $E_{g(LAO)}$ = 6.5~eV and ${d^{cr}_{\rm{LAO}}}$ = 18.5~u.c. from the main text, we obtain a value of $m^*$ = 0.14$m_o$ for the \lao ~effective mass. This value of $m^*$ is of the same order as the value of 0.27$m_o$ found in the literature for MOS capacitors fabricated using \lao ~as the dielectric.~\cite{chang2008}  The calculated curve for $J_Z$ vs. ${d_{\rm{LAO}}}$ is shown on the right in Fig.\ 1c in light blue, and is in excellent agreement with the data.   %This same equation was also used for the fits to the $JV$ curve shown in Fig.\ 1c. of the main text.

\begin{figure}
\includegraphics[width=\columnwidth]{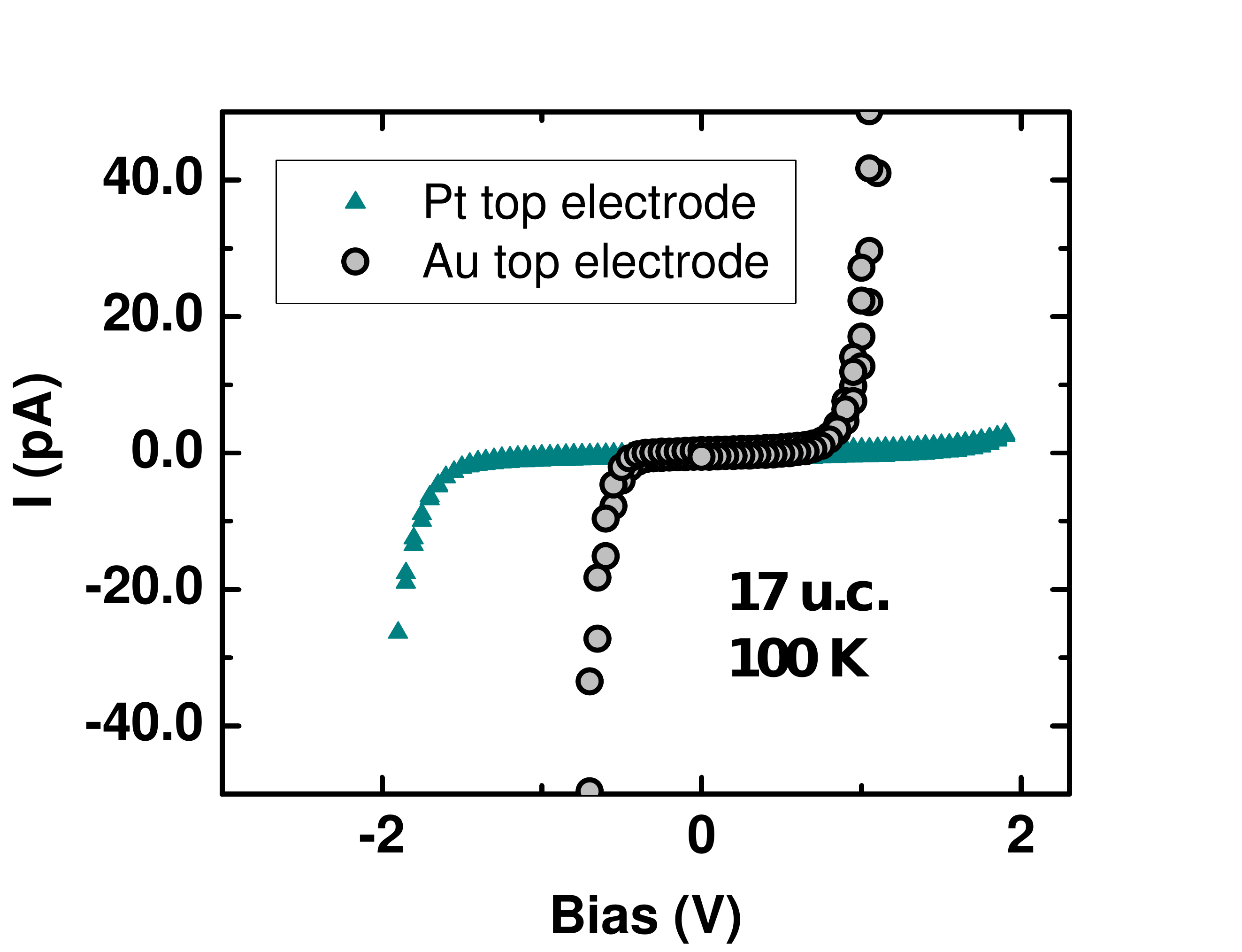}
\caption{$IV$ curves for a 17~u.c. sample using Pt and Au electrodes with the same area.  Higher currents are measured when using a metal (Au) with a lower work funtion than Pt.}
\label{f3}
\end{figure}

When tunneling from the \lao ~valence to \sto ~conduction band is considered instead, we have:
\begin{equation}
    \label{eq8}
\Phi(x)= E_{g(STO)}-qE_{bi}x+{\Delta}E_{V}+\epsilon .
\end{equation}
Using $E_{g(STO)}$ = 3.2~eV, ${\Delta}E_{V}$ = 0.35~eV and $E_{bi}$=($E_{g(STO)}+{\Delta}E_{V})/{d^{cr}_{\rm{LAO}}}$, we are unable to achieve agreement with the data shown in Fig.\ 1c of the main text.  Thus, we conclude that interband tunneling across \lao ~is the correct scenario.%We instead obtain a calculated curve which is on the same order of magnitude for the observed current with $m^*$ = 0.03$m_\circ$ and $E_{g(STO)}+{\Delta}E_{V}$ = 3.55~eV.  Since the value obtained for $m^*$ is considerably smaller than expected, and the calculated curve does not agree well with the data, our analysis suggests that Zener tunneling across \lao ~is a more likely scenario.
%
%The third possible scenario mentioned above is tunneling from the metal electrode into the \lao ~conduction band above a critical thickness that aligns the metal with the \lao ~conduction band.  The functional form for describing this type of tunneling is exactly the same as for Zener tunneling.  However, this case requires fermi level pinning at the \lao ~surface.

As noted earlier, we stress that regardless of the source of electrons during the tunneling process in Fig.\ 1c, the fact that a sudden increase in current density is observed above a critical thickness is indicative of a built-in electric field across \lao.  Our main conclusion is the same for either of the tunneling scenarios mentioned above.

We make one final note about the fits shown in Fig.\ 1c, main text:  All samples employ Pt electrodes while the 30~u.c. sample employs an Au electrode, as noted in the Methods section.  This may explain why $J$ is slightly higher than the theoretically calculated value for 30~u.c. in Fig.\ 1c, since Au has a lower work function than Pt.

\textbf{Metal work function dependence:}  As noted in the main text, depending on the relative value with respect to the \lao ~Fermi level, the metal work function will modify the net \lao ~built-in potential which adds to or subtracts from the ionic built-in potential.  In the presence of the electronic reconstructions, the exact band line-up is non-trivial and predictions require self-consistent calculation.~\cite{gu2009}  The difference in current densities between a tunnel junction with an Au vs. Pt electrode are shown in Supplementary Fig.~\ref{f3}.

\begin{figure}
\includegraphics[width=\columnwidth]{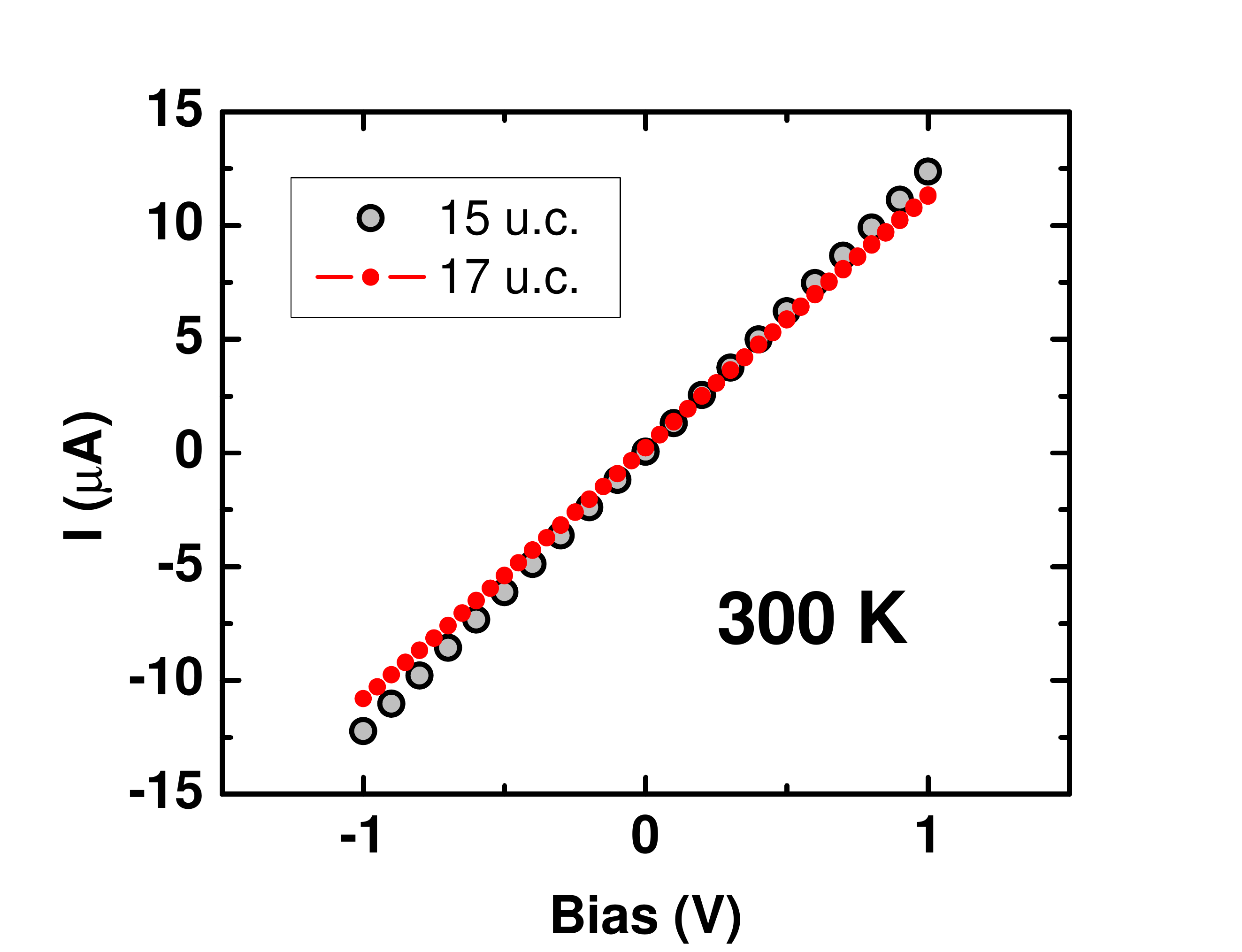}
\caption{$IV$ measured between two wire-bonded contacts on a 15~u.c.  and a 17~u.c. thick sample at room temperature.  Such linear $IV$ curves were measured on all samples to ensure metallic contacts to the electron gas.}
\label{f4}
\end{figure}

\textbf{Contact resistance:}  $JV$ curves between two contacts to the electron gas in each sample were measured to ensure linearity.  See Supplementary Fig.~\ref{f4} and compare to Supplementary Fig.~\ref{f3}.  This ensures that contacts are not the source of the observed rectifying $JV$.  

\textbf{Two-band model simulations:} Current voltage characteristics were calculated using the so called Keldysh formalism within the Non Equilibrium Green's Function approach (NEGF).~\cite{datta05} A two band Hamiltonian was set up where the band gap and effective mass of the conduction band and valence bands can be used as an input parameter. The advantage of using such a two-band model is the fact that any interband-tunneling process is automatically included. Finally a real space Hamiltonian was formed assuming nearest neighbor coupling.  To set up the electrostatic profile, a built in field $E_{bi}$ was assumed to be present across \lao.  Furthermore, a band offset $\Delta E_c$ was assumed at the \lao/\sto ~interface. 1D electronic transport was calculated by solving for the Green's function assuming infinite leads where the lead self-energies were calculated in a recursive fashion following a modified Sancho-Rubio approach.~\cite{sancho1984}  Finally, the the total current was calculated by summing over the modes in the transverse directions.

\begin{figure}
\includegraphics[width=\columnwidth]{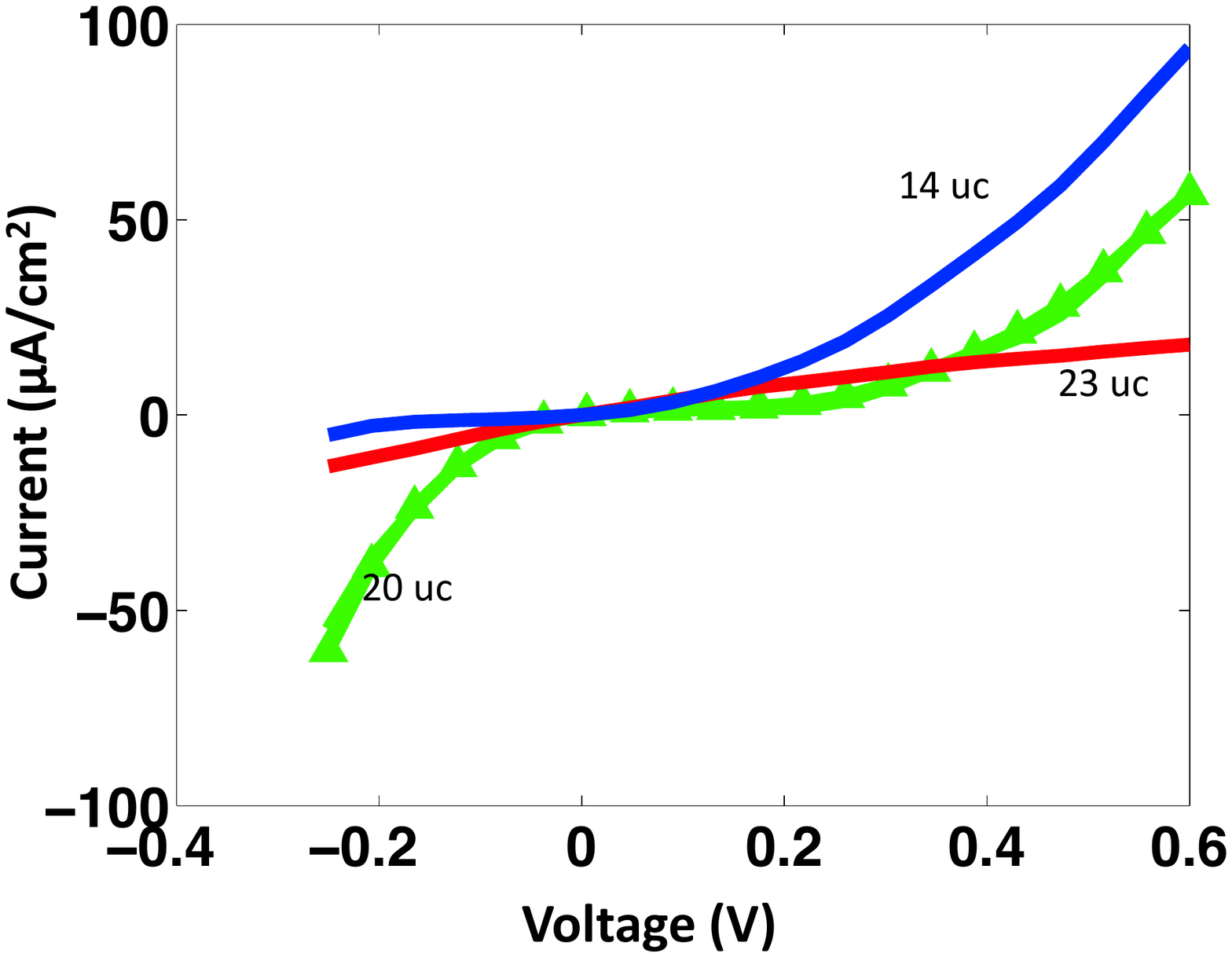}
\caption{Two band calculations reveal qualitative agreement with measured data shown in Fig.\ 1c of the main text:  1) The observed rectifying behavior is reproduced, 2) the same current value range is reproduced, and 3) the observed thickness dependence is reproduced.}
\label{f8}
\end{figure}

Fig. \ref{f8} shows the calculated $JV$ for various thicknesses of the LAO layer.  For this we have used a bandgap of $E_{g(LAO)}=6.5$ eV, a band offset $\Delta E_c=3.3$ eV and a built in electric field $E_{bi}\equiv0.9$~V/nm, corresponding to the band diagrams shown in Fig.\ 2 of the main text.  For consistency with the tunneling model fits, the effective mass was chosen to be $0.14m_0$.  The effective mass was kept unchanged through out the simulation domain for simplicity. In addition, since applied voltages are much smaller than the internal energy scales (e.g. band-gap, built-in voltage etc), charge self-consistency was ignored.

\begin{figure}
\includegraphics[width=\columnwidth]{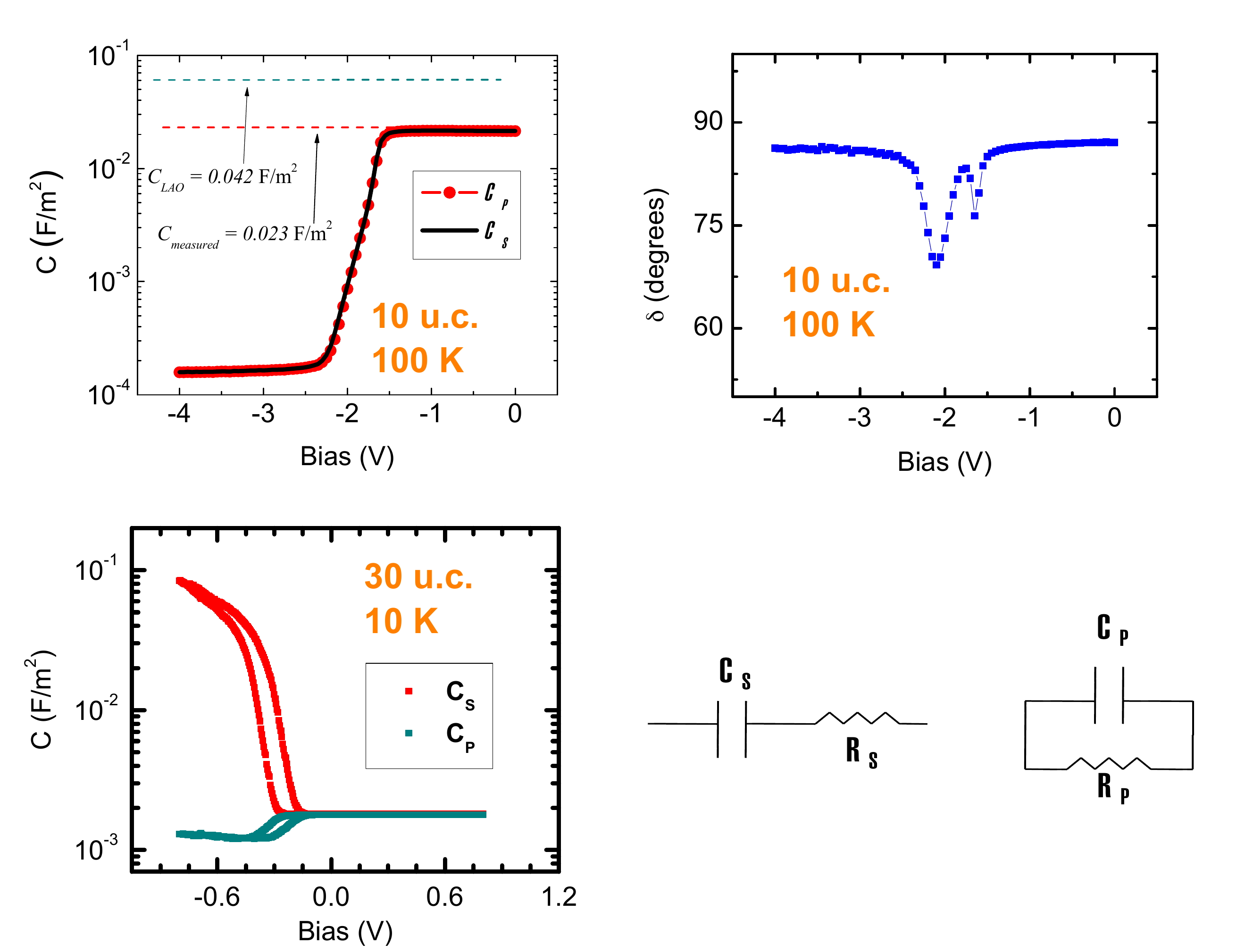}
\caption{\textbf{a,} shows both $C_P$ and $C_S$ for the 13~u.c. sample.  The ideal and measured capacitance across the \lao ~are labeled. \textbf{b,} Complex phase angle for the 13~u.c. sample. \textbf{c,} $C_P$ and $C_S$ for a 30~u.c. sample.  $C_P$ is also shown in Fig.\ 4a, main text.  \textbf{d} Circuit diagram for the parallel and series complex impedance model.}
\label{f7}
\end{figure}

As it can be seen, the calculated $JV$ show clear asymmetric behavior qualitatively reproducing the features obtained in the experiment. By looking at the local density of states we can verify that the current flows through direct tunneling in the forward bias ($+V$) while through an interband tunneling process in the reverse bias ($-V$). The results also reproduce the intriguing thickness dependence of the reverse bias ($-V$) current (with a maximum current density measured for the 20~u.c.). Although no particular effort was made to optimize the parameters to exactly fit the experiment, the chosen set of parameters reproduce the values of the current in the same quantitative range as obtained in the experiment. 

\textbf{Section II. Capacitance measurements across metal/L\lowercase{a}A\lowercase{l}O$_3$/S\lowercase{r}T\lowercase{i}O$_3$ ~tunnel junctions}

\begin{figure}
\includegraphics[width=\columnwidth]{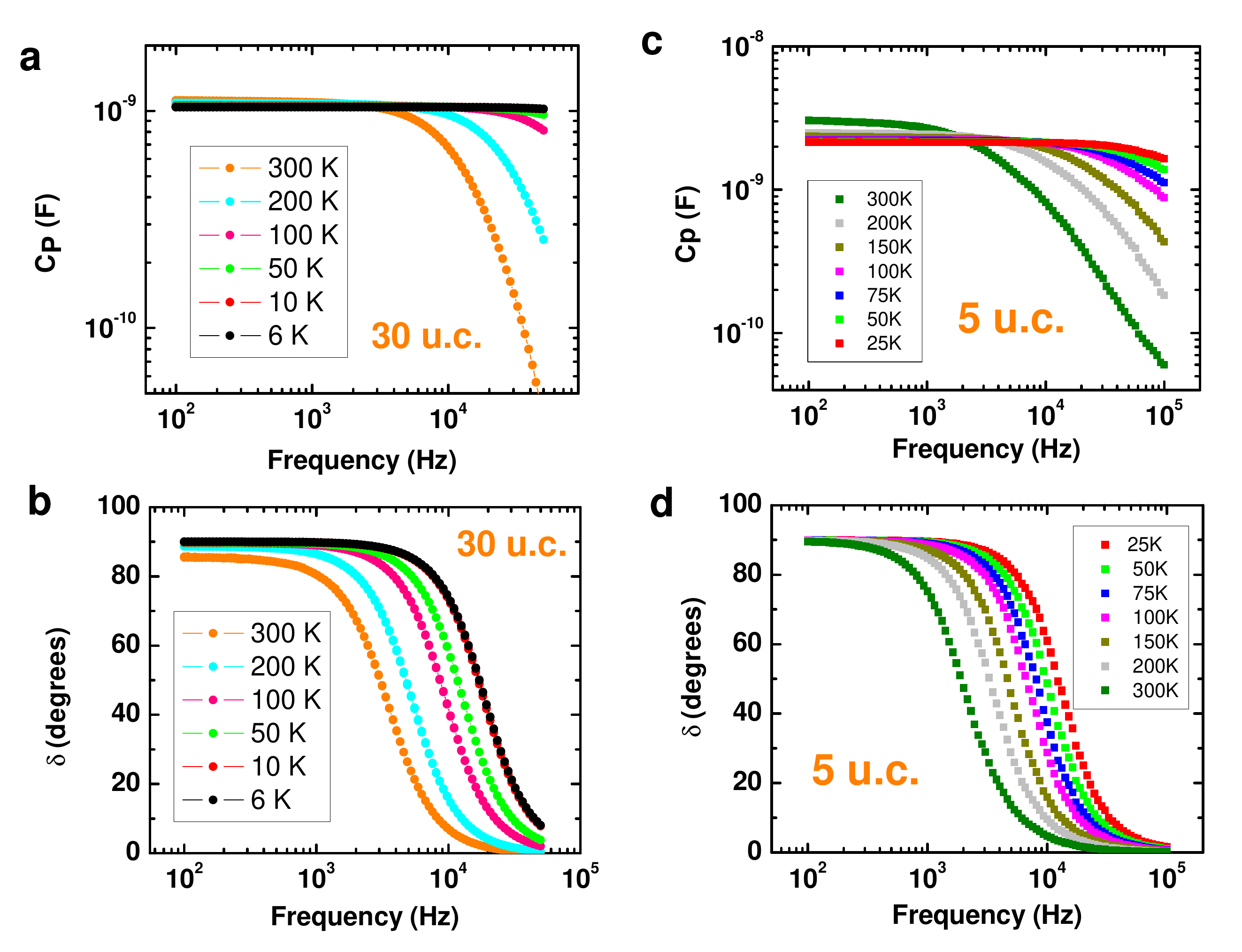}
\caption{$C_P$ vs. $f$ and $\delta$ vs. $f$ for both a 5~u.c. and a 30~u.c. sample showing the $RC$ roll-off frequencies.}
\label{f6}
\end{figure}

Capacitance measurements are often ambiguous and complicated by the fact that commercial capacitance bridges compare the measured complex impedance to either a series or parallel resistor-capacitor model shown at the bottom right of Supplementary Fig.\ \ref{f7}.  For an $ideal$ capacitor the series resistance (contact resistance) $R_S$ is negligible while the parallel resistance (dielectric resistance) $R_P$ is very large, and the measured capacitance is model independent (both $R_P$ and $R_S$ can be ignored).  Hence $C_P$ = $C_S$ for an ideal capacitor and the measured phase angle ($\delta$) of the complex impedance will be $\delta$ = 90$^{\circ}$.

Supplementary Fig.\ \ref{f7}a shows capacitance-voltage or $CV$ curves measured for a 13~u.c. sample, which is typical for samples with thicknesses below 20~u.c..  Supplementary Fig.\ \ref{f7}b  shows the complex phase angle, which remains at $\delta$ = 90$^{\circ}$, except for a peak during the depletion transition to the low $C$ value.  Thus, the sample exhibits near ideal capacitor response throughout the applied voltage range and is qualitatively similar to the response expected for a MOS capacitor.~\cite{sze2006}  The dashed lines indicate the ideal and measured capacitance expected across the \lao ~assuming bulk dielectric constant of $\approx$ 20.  The measured capacitance gives an \lao ~dielectric constant of $\approx$ 10, which is a factor of two less than expected.  This is however not unusual for ultra thin films since there are voltage drops associated with the electrode-dielectric interface which strongly affect the measured capacitance.~\cite{singhbhalla2003}

Supplementary Fig.\ \ref{f7}c shows a $CV$ curve for the same 30~u.c. sample shown in Fig.\ 4a of the main text.  The associated $\delta$($V$) curves are shown in Fig.\ 4 of the main text.  For samples with thicknesses $\geq$ 20~u.c., we find that $C_P$ and $C_S$ diverge while $\delta$ becomes less than 90$^{\circ}$, coinciding with the onset of Zener tunneling.  Thus as charge carriers are introduced into the \lao ~conduction band, the \lao ~effectively \textit{breaks down} and is no longer a good dielectric.  Thus we observe the sudden deviation in $\delta$ from 90$^{\circ}$ while the capacitance strongly depends on the circuit model chosen to analyze the data.  In the main text, we show $C_P$ in Fig.\ 4a since that is the value generally reported in the literature for MFIS or MFIM capacitors.
%In Supplementary Fig.\ XX, it can be seen that at the onset of tunneling, $C_P$ and $C_S$ diverge for samples with thickness $\geq$ 20~u.c..  This is true for both direct and zener tunneling.  Similar downturns in capacitance at the onset of tunneling have also been observed in metal-oxide-semiconductor (MOS) capacitors with ultra thin gate oxides~\cite{Zhang2001}. %For 20 u.c., we can safely assume zener breakdown.  For 40u.c., the onset of tunneling occurs when a high enough voltage causes significant bending and hence barrier thinning across the \lao.  The metal electrode is also at a higher potential than the \sto ~valence band at the onset of tunneling.  For thicker samples, the onsest of direct tunneling will happen at higher and higher applied voltages, since a larger bias will be required to achieve the same amount of band bending and barrier thinning across the \lao.  

The measurement frequencies for all data shown in Fig.\ 4, main text, were 100~Hz to 10~kHz.  As can be seen in Supplementary Fig.~\ref{f6} these are mostly below the RC roll-off limit at 10~K.
%(insert information about charge density, lao dielectric constant here, if necessary).
%Calculation of dipole moment in \sto.o
\bibliography{references3.bib}